\documentstyle[12pt,epsfig]{article}
\begin{document}

\begin{titlepage}
\rightline{April 2004}
\vskip 3cm
\centerline{\large \bf 
Supernova explosions, 511 keV photons, gamma ray bursts 
}
\vskip 0.3cm \centerline{\large \bf
and mirror matter}

\vskip 2.2cm
\centerline{R.~Foot$^{a}$ and Z.~K.~Silagadze$^{b}$~\footnote{
E-mail address: foot@physics.unimelb.edu.au, \hspace{1mm}
silagadze@inp.nsk.su}
}

\vskip 0.7cm
\centerline{\it $^a$ School of Physics,}
\centerline{\it University of Melbourne,}
\centerline{\it Victoria 3010 Australia}
\vskip 0.5cm
\centerline{\it $^b$ Budker Institute of Nuclear Physics,}
\centerline{\it  630 090, Novosibirsk, Russia}
\vskip 1cm
\noindent
There are three astroparticle physics puzzles which fire the imagination:
the origin of the ``Great Positron Producer'' in the galactic bulge, the
nature of the gamma-ray bursts central engine and the mechanism of supernova
explosions. We show that the mirror matter model has the potential
to solve all three of these puzzles in one beautifully simple strike.

\end{titlepage}

Recent observation of a 511 keV annihilation 
line by the SPI spectrometer on the 
INTEGRAL satellite, indicating
a powerful source of positrons in the central regions of the Galaxy, 
has spurred
a great deal of activity and imagination among astroparticle physics 
community. This is not surprising. In our matter dominated universe any
antimatter source of that magnitude certainly is extraordinary and deserves
close attention. What is surprising is why this has not happened earlier,
because the 511 keV photon emission from the galactic bulge was first detected
more than 30 years ago\cite{first}. Maybe the explanation is given by the 
Andre Gide's aphorism: ``Everything has been said before, but since nobody 
listens we have to keep going back and beginning all over again.''

The  INTEGRAL-SPI measured flux \cite{exp} confirms earlier findings and
equals
\begin{eqnarray}
{\rm flux} \approx  10^{-3} \ {\rm photons~cm^{-2}~s^{-1}}.
\end{eqnarray}
Besides, these new measurements provide crucial information on the geometry
of the source and firmly establish that the bulk of annihilation takes place
in the Galactic bulge at the distance of about 8~kpc \cite{Hyper}. 
The flux suggests a positron annihilation rate of
\begin{eqnarray}
R \approx 1.3\times 10^{43} \ {\rm s}^{-1},
\label{rate}
\end{eqnarray}
if it is assumed that the annihilation proceeds via positronium and the 
positronium fraction is 0.93, as implied by observations \cite{Hyper}.

The origin of the positrons is unclear. Certainly a conventional
explanation might be possible, 
and several have been proposed (see e.g.\cite{conv}).
More exotic mechanisms are also possible, including the recent
suggestion of decays or annihilations of hypothetical dark matter particles in
the MeV mass range\cite{mevp,mevp2}.
In this note, we would like to point out that a natural
mechanism for positron production occurs in the mirror matter
model and that this mechanism can help supernovas to
explode and also explain gamma ray bursts. 

Recall, mirror matter is predicted to exist if nature exhibits
an exact unbroken mirror (exact parity) symmetry\cite{flv}. The general idea, 
as well as some details, have been reviewed
many times (see e.g. Ref.\cite{review}). 
But since, perhaps, only a small part of the astroparticle physics community
listens, we will go back and repeat the main points again.  

From modern perspectives, symmetry principles play a crucial role in the Grand
Design of the Universe, space-time and gauge symmetries being the most 
important players. The proper Lorentz group is isomorphic to
the $SL(2,C)$ group 
and this enables one to define notions of left and
right chirality at the fundamental level of spinor fields. 
One can expect that the
gauge symmetry group $G$ (where $G = SU(3) \otimes SU(2) \otimes U(1)$ is the 
simplest case) treats the left and right chiral fields on an
equal footing. 
Surprisingly this is not the case: weak interactions are left-handed and
no right-handed neutrinos have been observed. 
Perhaps the best explanation of this
puzzle is given by the anthropic principle.

One needs neutrinos and weak interactions to ensure our existence, because
they play an important role in the Sun's energetics. But neutrinos cannot be 
too
massive otherwise they will overclose the Universe \cite{zeld}
and make the existence of intelligent observers impossible. In fact this 
cosmological bound implies \cite{nmass} $\sum m_\nu \le 40~{\mathrm eV}$. 
In the presence of right-handed neutrinos nothing will prevent neutrinos to 
acquire Dirac masses of the order of the accompanying lepton masses. 
Therefore the
Designer of our universe is bound to violate parity invariance in order 
for the
universe to be hospitable to observers\footnote{
One consequence of this is that the right-handed neutrinos can
be gauge singlets which means that they are able to gain a large
Majorana mass, leading to small effective $\nu_L$ masses (see-saw
mechanism).}.

In spite of the apparent parity non-invariance of the ordinary
particles, the universe could still be left-right
symmetric if {\bf CP} were an exact symmetry\cite{landau}. 
But this option is not allowed
too, because in the {\bf CP}-symmetric universe there will be no baryonic
asymmetry and hence no observers \cite{sakh} (it is also
ruled out by experiments on kaons and B-mesons!).

However, there is a subtle way out. It is
possible to build a left-right symmetric universe by 
doubling the gauge symmetry so that the full gauge group 
becomes $G \otimes G$. 
Then for each type 
of ordinary particle (electron, quark, photon etc) there is a mirror partner
(mirror electron, mirror quark, mirror photon etc), 
of the same mass. The two sets of particles form 
parallel sectors each with gauge symmetry $G$ -- except that where
the ordinary particles have left-handed (V-A) weak interactions,
the mirror particles have right-handed (V+A) weak interactions.
The unbroken mirror symmetry maps
ordinary particles into mirror
particles and includes space inversion
(the explicit transformation is given in Ref.\cite{flv}). 
Exact unbroken time reversal symmetry
also exists in this model \cite{flv}.
Therefore the full Poincar\'{e} group, including improper Lorentz 
transformations, becomes an exact symmetry group of nature. This is certainly
appealing esthetically. However, an even more subtle point is the
realization
that this restoration of left-right symmetry might be necessary for anthropic 
reasons, just like the previous two 
steps ({\bf P} and {\bf CP} violations). Justification
comes from considering possible interactions between two parallel sectors
as we will now explain.

Ordinary and mirror particles should interact with each
other by gravity.  Some other interactions are also allowed\footnote{Given the 
constraints of gauge invariance, renomalizability and mirror
symmetry it turns out\cite{flv} 
that the only allowed non-gravitational interactions
connecting the ordinary particles with the mirror particles
are via photon-mirror photon kinetic mixing 
and via a Higgs-mirror Higgs quartic
coupling, ${\cal{L}} = \lambda \phi^{\dagger} \phi \phi'^{\dagger}
\phi'$. If neutrinos have mass, then ordinary - mirror
neutrino oscillations may also occur \cite{flv2}.}
and for our goals the most important one is the possible photon-mirror
photon kinetic mixing interaction:
\begin{eqnarray}
{\cal{L}} = {\epsilon \over 2} F^{\mu \nu} F'_{\mu \nu}, 
\label{km}
\end{eqnarray}
where $F^{\mu \nu}$
($F'_{\mu \nu}$) is the field strength tensor for electromagnetism
(mirror electromagnetism). 
Photon-mirror photon mixing causes 
mirror charged particles to couple to 
ordinary photons with a small effective
electric charge, $\epsilon e$\cite{flv,hol,sasha}.

Mirror matter is necessarily dark and stable and is an impressive
dark matter candidate which has been extensively
studied in recent years\cite{dm}.
Hard experimental evidence for mirror matter-type dark
matter comes 
from the impressive DAMA/NaI
experiment\cite{dama}.
It turns out that this experiment is consistent with
halo dark matter consisting of mirror particles interacting
with ordinary particles via the photon-mirror
photon kinetic mixing interaction, Eq.(\ref{km})\cite{f1}.
A fit to the DAMA annual modulation signal suggests\cite{f1}
\begin{eqnarray}
\epsilon \sim 5\times 10^{-9}
\end{eqnarray}
Importantly, the mirror matter explanation of the DAMA
signal is not in conflict with any other experiment\cite{f1,f2}.

We now consider the effect of photon-mirror photon kinetic
mixing, Eq.(\ref{km}), in the core of an {\it ordinary} type II 
supernova (the more interesting case of a mirror
supernova will be considered in a moment).
The main production process for minicharged particles 
in the core of a mirror supernova is the plasmon decay process (see
e.g. \cite{raffelt} for a review).
The energy loss rate for production of minicharged
particles has been calculated in Ref.\cite{raffelt}:
\begin{eqnarray}
Q_p = {8\zeta(3) \over 9\pi^3} \epsilon^2 \alpha^2 \left(
\mu_e^2 + {\pi^2T^2 \over 3}\right) T^3 Q_1
\label{Qp}
\end{eqnarray}
where $Q_1$ is a factor of order unity. $Q_p$ exceeds the
energy loss rate due to neutrino emission for
$|\epsilon | \stackrel{>}{\sim} 10^{-9}$\cite{raffelt}.
This of course assumes that the minicharged particles freely
stream out of the core. This will not be the case for
$e'^{\pm},\gamma'$ (we denote mirror particles with a prime) 
because they will be trapped by their own mirror electromagnetic
interactions. So for $\epsilon \sim 5 \times 10^{-9}$, we expect 
$e'^{\pm}, \gamma'$ to roughly thermalize with the ordinary particles
in the core of the supernova. 
This means that the energy loss rate due to
emission
of $e'^{\pm}, \gamma'$ need not be 
much greater than that due to neutrino emission (however
it should be comparable).
Of course, direct detection of $e'^{\pm}, \gamma'$ from a supernova 
seems to be difficult if not impossible for ordinary
matter observers. 

Let us stress that $\epsilon \sim 5 \times 10^{-9}$, implied by the DAMA 
annual modulation signal \cite{f1}, is just in the interesting range where
the effects of the photon-mirror photon kinetic mixing 
becomes important for supernova
energetics and dynamics. For $\epsilon$ of that magnitude the supernova 
energy loss through mirror channels becomes comparable to that due to 
neutrino emission. This raises an interesting question.

It is believed that neutrinos drive the evolution of the collapsing supernova
core because they dominate the event energetically. Supernova explosions of
massive stars are explained by convectivelly supported neutrino-heating
mechanism \cite{collaps}. But recent refined simulations showed \cite{supern} 
that there is insufficient neutrino energy transfer behind the
stalled supernova shock to produce the explosion. This suggests some missing 
physics. In light of what was said above, we suggest that this missing
physics is provided by the photon-mirror photon kinetic mixing:
The $e'^{\pm}, \gamma'$ produced in the core will interact
and heat the matter behind the shock (adding to the effect
of neutrino-heating) which might help 
produce the explosion. Let us see if this is reasonable.
The cross section for
MeV $\gamma'$ (and large angle $e'^{\pm}$) scattering
with ordinary electrons and positrons (i.e. $\gamma' + e^{\pm} 
\to \gamma + e^{\pm}$ and $e'^{\pm} + e^{\pm} \to e'^{\pm} + e^{\pm}$)
is of order 
\begin{eqnarray}
\sigma \sim \epsilon^2 \pi r_0^2 \sim 
10^{-41}\left({\epsilon \over 5\times 10^{-9}}\right)^2 \ {\rm cm}^2 
\end{eqnarray}
where $r_0 = \alpha/m_e$ is the
classical radius of the electron. Remarkably this is roughly the
same size as the neutrino nucleon cross section,
\begin{eqnarray}
\sigma (\bar \nu_e p \to n e^+) &=&
{4G_F^2 E_\nu^2 \over \pi}
\nonumber \\
&\approx &
10^{-41} \left({E_\nu \over 10\ {\rm MeV}}\right)^2 \ {\rm cm}^2
\end{eqnarray}
where $E_\nu$ is the energy of the neutrino. Importantly, the
energy dependence is different:
compared with neutrino interactions,
the mirror particle interactions with ordinary matter
are larger at lower energies\cite{detail}.
Evidently the heating effect of the mirror particle interactions on
the ordinary matter just behind the shock is expected
to be comparable to -- or may even exceed -- the neutrino effect. 
This seems to be
rather nice for supernova explosions!

It appears, therefore, that the anthropic argument outlined above becomes 
beautifully complete. One needs supernovas to explode to make heavy elements
which are crucial for our existence. But in a universe without mirror 
particles supernovas do not explode, therefore such universe is devoid of
observers. The Designer {\it has} to restore the left-right symmetry by
introducing mirror particles and arrange the photon-mirror photon
kinetic mixing
of the right magnitude to ensure the appearance of intelligent observers. 

In the case of a {\it mirror} type II supernova, things are no less
interesting. In this case, the core of the mirror 
supernova would be a source of ordinary electrons, positrons
and gamma rays -- making such an event easily detectable
for ordinary matter observers. 
During a mirror supernova
explosion an enormous energy will be deposited in a short time and a
small volume in mildly relativistic $e^+e^-\gamma$ plasma. The resulting
fireball will lead to a gamma ray burst (GRB) \cite{GRB} 
provided that the number of ordinary baryons in the 
mirror supernova is sufficiently low. In fact the gamma
ray burst has roughly
the right characteristics (energy release, time scale, and
potentially small baryon load)
to be identified with the observed gamma ray bursts as pointed out by 
Blinnikov\cite{blin}\footnote{
Blinnikov considered neutrino-mirror neutrino oscillations
(rather than the photon-mirror photon kinetic mixing
interaction) as the mechanism to convert mirror particles into
ordinary particles in the core of a mirror supernova.
However, Volkas and Wong\cite{wong} showed that that
mechanism was not viable due to matter effects which
suppress neutrino-mirror neutrino oscillations. }.
In this case one might expect that the
distribution of GRBs
should be correlated with the
distribution of dark matter 
which is not excluded \cite{blin1}. 

For the purposes of this paper we will 
simply define $f_{e^+}, f_{e^-}, f_{\gamma}$ as the fraction of
the total energy of the collapsing mirror star released 
as $e^+, e^-, \gamma$ respectively (of typical energy of order
10 MeV).
We will not attempt to precisely calculate $f_i$, since the details are quite
complicated with many sources of uncertainty. Crudely speaking, 
$f_i$ would be expected to be of order 0.1.

The total energy released in mirror supernova should be
similar to ordinary supernova which is of order
\begin{eqnarray}
E_{SN} \sim 10^{53-54} \ {\rm ergs}.
\end{eqnarray}
This means that the number of $e^{\pm}, \gamma$ produced
in a mirror type II supernova should be roughly:
\begin{eqnarray}
N_{e^+} \simeq N_{e^-} \sim N_{\gamma}
&=& {f_i E_{SN} \over \langle E_i \rangle}
\nonumber \\
&\sim & 10^{57}
\left( {f_i \over 0.1}\right) 
\label{Npos}
\end{eqnarray}
where $i$ labels $e^+, e^-, \gamma$
and we have used $\langle E_i \rangle \sim 10$ MeV.

The positrons (and electrons) will be confined by the magnetic field
of the galaxy
and eventually thermalize and annihilate
(with stopping distance of order $10^{25}$ cm\cite{mevp}
and time scale $\stackrel{>}{\sim} 10^8$ years)
producing two 511 keV gamma rays.
Thus, the 511 keV gamma ray production rate is
\begin{eqnarray}
{d N_{\gamma} \over dt} \sim \left({ f_{e^+} \over 0.1}\right) 10^{57} R_{SN}
\end{eqnarray}
where $R_{SN}$ is the rate of mirror supernova explosions with
sufficiently small ordinary baryon load in the 
galactic bulge. Equating this rate to the measured rate,
Eq.(\ref{rate}), gives a rate of roughly,
\begin{eqnarray}
R_{SN} \sim 10^{-2} \ {\rm per\ million\ years}
\end{eqnarray}
Of course this is only an order of magnitude estimate
with many sources of uncertainty. For example, it may
turn out that a significant fraction of
positrons annihilate before becoming non-relativisitic, which
would mean that $R_{SN}$ would be somewhat larger
than our estimate above. Anyway,
according to the estimate in Ref.\cite{mevp}, the positrons should
be confined (by the magnetic field) to a region of order a parsec from
their production point. Thus, we would not expect a
uniform positron emission from the bulge, but rather a finite set
of sources, each in the vicinity of a mirror type II supernova.
This is one way, in principle, to distinguish this 
explanation from explanations
involving exotic MeV particle decays or annihilations\cite{mevp,mevp2}.
The observations are consistent with a uniform distribution
but future observations may reveal some substructure.

Interestingly, the rate, $R_{SN}$, inferred above is also roughly
consistent with the observed rate of gamma ray bursts:
a mirror supernova rate of $10^{-2}$ per million years in our galaxy,
would suggest a rate of about $10^2$ per year per 10 billion galaxies
(of order the number of galaxies in the observable universe). This
is roughly the observed rate of gamma ray bursts.
Of course there is no reason for this rate to be exactly
uniform -- so this extrapolation cannot be rigorous.

Note that $R_{SN}$ is much less than the rate of ordinary
type II supernova explosions in our galaxy.
This probably means that most mirror type II supernova
do not lead to gamma ray bursts. This might occur if
most mirror supernova have enough ordinary baryons to
absorb the energy of the expanding $e^{\pm},\gamma$ fireball.
Even a relatively small proportion of ordinary baryons
($\sim 10^{-5} M_{\odot}$) might be enough, since
one must take into account the effect of proton-mirror
proton collisions which could rapidly cool the protons,
dumping energy into the collapsing mirror star and potentially
also help power the explosion.
In this way a small baryon load might act as a type of
catalyst allowing energy from the relatistic $e^{\pm},\gamma$
fireball to be converted into heating the mirror baryons.
Very small baryon load ($\ll 10^{-5} M_{\odot}$) is also 
problematic for observations.
Associated clean fireballs produce \cite{load} intense, extremely  short
(subsecond) gamma-ray transients with very energetic ($\gg 10~\mathrm{MeV}$)
gamma-rays. Such events are hard to detect because of dead time and
sensitivity limitations of previous gamma-ray detectors \cite{load}.

Finally, note that idea that the 511 keV annihilation radiation might be
connected with gamma ray bursts is not new.
In particular, Ref.\cite{purcell} considered the possibility
of a gamma ray burst occurring in the galactic center
within the past $10^6$ yr might explain the 511 keV
photon emission.
Irrespective of the nature of the GRB central engine, copious pair production
is expected during fireball bursting phase due to $\gamma-\gamma$ absorption
between high energy photons with estimated number of produced pairs
$\sim 10^{54} E_\gamma^2$, where $E_\gamma$ is the GRB energy released in
gamma-rays in $10^{52}~\mathrm{ergs}$ units \cite{pair}. This estimation
does not depend on any special model, but only assumes that the observed
intrinsic GRB spectrum can be extrapolated
to very high energies \cite{pair}. Nevetheless it is close to our rough
estimation Eq.(\ref{Npos}). Even more important is that the new born $e^{\pm}$ 
pairs survive and do not annihilate with each other into $\gamma$ rays again 
because their annihilation time for
typical GRB parameters is much longer than the hydrodynamic time in the
comoving frame \cite{pair}.

In conclusion, we have shown that mirror matter
provides a simple physical picture in which to
explain three
astroparticle puzzles: origin of positrons in
the galactic center, the nature of the gamma-ray bursts
central engine and the mechanism of supernova explosions.
This is of course, just a beginning  -- more detailed
calculations would be helpful, along with additional
experiments to more precisely pin down the parameter,
$\epsilon$, which couples the ordinary world to the mirror
one. However, the fact that the mirror particles are expected to be
generated 
in a supernova core and their cross sections with ordinary matter turns out to 
be about the same as the neutrino one (for energies
relevant for supernovas) gives significant encouragement 
that we are on the right track: if the cross section turned out
to be much larger than the neutrino one
then the $e^{\pm},\gamma$ would never escape the mirror supernova which
means that we
could not explain 
gamma ray bursts or galactic positrons, if the cross section turned
out to be much smaller, 
it could not help supernovas to explode. Of course,
we need supernovas to explode, not just to agree
with observations, but more importantly, to generate
heavy elements. From this anthropic point of view,  
gamma-ray bursts and galactic bulge
positrons could be viewed as simply byproducts of our own existence! 


\vskip 0.5cm
\noindent
{\large \bf Acknowlegements}
\vskip 0.3cm
\noindent
The authors would like to thank S. Blinnikov for valuable comments
on a draft of this paper.


\begin{thebibliography}{999}
\bibitem{detail}
The energy dependence of the cross section for
mirror particle interactions with ordinary matter is
quite different to neutrino interactions since
the former is an electromagnetic interaction while the
latter is a weak interaction. Considering the
process $\gamma' e^{\pm} \to \gamma e^{\pm}$, the 
total cross secton is
$$
\sigma = \epsilon^2 \pi r_0^2 \left[y(1 -2y -2y^2) \ln{ (1 + {2 \over y})} 
+ 4y^2 + {2y(1+y) \over (y + 2)^2}\right]
$$
where $y = m_e/E_{\gamma'}$. This cross section has 
the following limits:
$\sigma = 8\epsilon^2 \pi r_0^2/3 \simeq 2\times 10^{-41}\left( {\epsilon
\over 5\times 10^{-9}}\right)^2 \ {\rm cm}^2 \ \ {\rm for}\ y \gg 1
\ ({\rm Thomson \ limit}) 
$
and 
$ \sigma = \epsilon^2 \pi r_0^2 \left[ y\left (\ln{ {2 \over y}}+ 
{1 \over 2} \right ) \right]
\ \ {\rm for}\ y \ll 1.
$
In other words, the cross section becomes energy independent
at low energies $E_{\gamma'} \ll$ MeV and slowly falls
at high energies $E_{\gamma'} \gg$ MeV.
\end{thebibliography}
\end{document}